\documentstyle[12pt]{article}
\begin{document}
\baselineskip = 20pt
\input epsf
\ifx\epsfbox\UnDeFiNeD\message{(NO epsf.tex, FIGURES WILL BE IGNORED)}
\def\figin#1{\vskip2in}
\else\message{(FIGURES WILL BE INCLUDED)}\def\figin#1{#1}\fi
\def\footnotefont{\tenpoint}

\parindent 25pt
\overfullrule=0pt
\tolerance=10000
\def\Re{\rm Re}
\def\Im{\rm Im}
\def\titlestyle#1{\par\begingroup \interlinepenalty=9999
     \fourteenpoint
   \noindent #1\par\endgroup }
\def\tr{{\rm tr}}
\def\Tr{{\rm Tr}}
\def\half{{\textstyle {1 \over 2}}}
\def\calt{{\cal T}}
\def\ie{{\it i.e.}}
\def\np{Nucl. Phys.}
\def\pl{Phys. Lett.}
\def\pr{Phys. Rev.}
\def\prl{Phys. Rev. Lett.}
\def\cmp{Comm. Math. Phys.}
\def\quart{{\textstyle {1 \over 4}}}
\def\RR{${\rm R}\otimes{\rm R}~$}
\def\NSNS{${\rm NS}\otimes{\rm NS}~$}
\def\RNS{${\rm R}\otimes{\rm NS}~$}
\def\calf{${\cal F}$}

\baselineskip=14pt
\pagestyle{empty}
{\hfill DAMTP/96-110}
\vskip 0.1cm
{\hfill hep-th/9612127}
\vskip 0.4cm
\centerline{CONFIGURATIONS OF TWO D-INSTANTONS.}
\vskip 1cm
 \centerline{ Michael B.  Green\footnote{M.B.Green@damtp.cam.ac.uk}
 and Michael Gutperle\footnote{M.Gutperle@damtp.cam.ac.uk}}
\vskip 0.3cm
\centerline{DAMTP, Silver Street,}
\centerline{ Cambridge CB3 9EW, UK.}
\vskip 1.4cm
\centerline{ABSTRACT}
\vskip 0.3cm

 The potential between two separated  D-instantons at fixed (super)  
space-time
points is obtained  by a simple explicit integration over  the \lq  
massive'
variables of the zero-dimensional reduction of  ten-dimensional  
$U(2)$ super
Yang--Mills theory.     This potential vanishes for asymptotically large
separations, becoming significant at separations of around  the  
ten-dimensional
Planck scale  with a singularity at the   origin, which is resolved  
by the
extra \lq massless'  internal Yang--Mills super-coordinates.
\vfill\eject
\pagestyle{plain}
\setcounter{page}{1}

The emerging  unified description of  superstring theory and  
eleven-dimensional
supergravity is largely based on  appreciation of the $p$-brane solitonic
solutions of these theories and their compactification to lower  
dimensions.
The  D-brane description of  stringy $p$-branes that carry  \RR charges
\cite{polchinski} has given a well-controlled perturbative  
formulation of this
class of stringy soliton and has lead to a magnificent set of  
insights into
stringy effects, black hole quantum mechanics and properties of  
supersymmetric
quantum field theory.
 Compactification of   euclidean $p$-brane world-volumes leads to  
instanton
configurations that  also play   significant r\^oles in the  
complete theory
(as explored in, for example,
\cite{becker,ooguri,harveymoore,witteninstantons}.

In addition to the solitonic branes, type IIB superstring theory  
possesses an
instanton solution in ten dimensions   --  the  $p=-1$ D-brane   
that couples
to the pseudoscalar \RR charge.  Such instantons presumably have  
important
effects in the theory -- for example, they are responsible for  \lq  
point-like'
effects in fixed-angle scattering.    Indeed,  the possibility of  inducing
point-like structure was one of the original reasons for studying string
theories with Dirichlet boundary conditions in both the bosonic theory
\cite{greenone,gutperle} and the superstring \cite{greeniib}.
 In a separate paper \cite{gutnew}  we  discuss such point-like  
scattering
in the background of a D-instanton and also determine certain  
non-perturbative
terms that  are induced  in the effective potential by the D-instanton
background.

Here we will focus on multi D-instanton configurations in uncompactified
ten-dimensional space-time.   The classical D-instanton solutions of IIB
supergravity theory in \cite{gibbons} have the interpretation, in  
the string
frame, of Einstein--Rosen space-time wormholes.   Some integer unit  
of  \RR
pseudoscalar flux disappears through any given wormhole into an  
unphysical
universe.  Thus, the sector of the moduli space with charge $N$ has  
$p(N)$
wormholes ending on distinct unphysical universes, where $p(N)$ is the
partition of $N$ into integers.      The classical description is  
obviously
inadequate  since the string coupling gets large in the wormhole  
neck at around
the Planck scale.   To go further we  need to analyze the stringy  
description
of  the D-instanton configuration space.

 A configuration of separated  D-instantons  is described by  
world-sheets with
boundaries fixed at space-time points representing the location of the
instantons \cite{polchcom,greentwo}.  Such world-sheets can be  
described by
gluing open-string strips together, where  the open \lq strings' have
end-points fixed in {\it space-time}.   This is conformally  
equivalent to the
insertion of point-like densities on closed strings  
\cite{greenone}.    The sum
over  world-sheets with fixed positions for the instantons is  
equivalent to
integrating over the \lq stretched open strings' joining them.
However,  when D-instantons  coincide
there are enhanced symmetries and these integrations are singular.

The moduli space of  $n$ D-branes  with world-volumes of
dimension $p+1$ is described by the  reduction   to  $p+1$ dimensions of
ten-dimensional supersymmetric $U(n)$ Yang--Mills theory  
\cite{witten}.  In
other words the Yang--Mills potential, $A_{A\mu}$,  and the  
Majorana--Weyl
fermion, $\psi_A$, are taken to be independent of the $9-p$ dimensions
transverse to the world-volume.\footnote{Our conventions are that  
the index $A$
denotes the adjoint of $U(n)$ and takes $n^2$ values,  
$\mu=0,1,\cdots,9$ is a
$SO(9,1)$ vector index and $a  = 1,\cdots,16$ is a $SO(9,1)$ Weyl spinor
index.}  The D-instanton configuration space is determined by the  
$p=-1$ case.
This  is a model of  $n\times n$ bosonic and fermionic $U(n)$
matrices,
$A_\mu=A_{A\mu} T^A$ and $\psi=\psi_AT^A$, where $T^A$ are the
generators of the Lie algebra of $SU(N)$ satisfying
$[T^A,T^B]=f^{ABC}T^C$ with $f^{ABC}$ being the structure
constants. In addition there is  a   $U(1)$  which
describes the overall centre of mass degrees of freedom -- since
all variables
are in the adjoint of the group, they are all uncharged under this
$U(1)$.   The $D$-instantons are solutions of the euclidean theory so the
Minkowski signature   fields should be Wick rotated in order to give a
well-defined measure.    In the latter part of the paper we will  
adopt the
procedure of carrying out as much of the calculation as possible  
with Minkowski
signature in order to make use of well-known manipulations of $SO(9,1)$
Majorana--Weyl  fermions and  Dirac $\Gamma$ matrices.

The D-instanton action  is given by simply  deleting all  
derivatives in the
supersymmetric Yang--Mills action,
\begin{equation}\label{dinstaction}
S = {1\over 2  \kappa }\left({1\over 4}  \tr([A_\mu,A_\nu]^2)+{i\over 2}
\tr(\bar{\psi}\Gamma^\mu
[A_\mu,\psi])\right) ,
\end{equation}
where $\kappa$ is the string coupling constant.
The minimum potential   of the system is given by setting the
fermions to zero
and $[A_\mu,A_\nu]=0$. These conditions are satisfied by matrices  
$A_\mu$ in
the $U(1)^n$ Cartan subalgebra  which have the form,
\begin{equation}
  A_\mu  =\left(\begin{array}{cccc}Y_1^\mu&&\\
&Y_2^\mu&\\
& &..&\\
& && Y_n^\mu
\end{array}\right),
\end{equation}
where $I=1,\dots,n$ and the $Y_r^\mu $  are the positions of  $n$   
separated
D-instantons.
The partition function  is simply an integral of $e^{-S}$ over
bosonic and fermionic matrices.    The measure on the reduced  
moduli space of
$n$ separated instantons is obtained by integrating all variables  
that are not
in the Cartan subalgebra, so that the partition function can be  
expressed as,
\begin{equation}
\label{partfun}
{\cal Z}^{(n)}= \int \prod_{I=1}^{n} d\psi_I  dA_I Z^{(n)  } [\psi_I,A_I]
e^{J^I \psi_I},
\end{equation}
where $J^I$ represent closed-string sources and
\begin{equation}
\label{primedef}
 Z^{(n)} [\psi_I,A_I] = \int  \prod_{A'=1}^{ n^2-n} d\psi_{A'}   
dA_{A'} \  \exp
\left( -S(A,\psi) \right) ,
\end{equation}
and  ${}_{A^{\prime}}$ indicates that the elements in the Cartan  
subalgebra are
not included.

The measure $Z^{(n)}$ is trivially independent of the variables in  
the overall
diagonal $U(1)$ so that the partition function  (\ref{partfun})  
includes a
volume  factor  due to integration over the corresponding  ten  
bosonic and
sixteen fermionic variables  (which are identified with the sixteen  
broken
supersymmetries).   We are here thinking of soaking up the overall  
fermionic
variables with external sources which enter in the calculation of  
closed-string
scattering amplitudes \cite{gutnew}.  We want  to evaluate the  
measure for the
remaining integrations in (\ref{partfun}), which is given by the  
integration
over the \lq internal variables' in (\ref{primedef}).  This  
corresponds to the
situation in which the D-instantons are all separated and we are  
integrating
over the \lq stretched strings' joining them.

We will now specialize to the case of two D-instantons, for which  
the relative
moduli space is determined by the group $SU(2)$.
 In order to simplify the expression it  will later prove very   
convenient to
make use of symmetries of the action  (\ref{dinstaction}) in order  
to pick a
particular parameterization of the integration variables.
One of these symmetries is the remnant of the local $SU(2)$ gauge  
symmetry,
with parameter $\Lambda^A$ ($A=1,2,3$).  In addition, the action is  
invariant
under   $SO(9,1)$ Lorentz
transformations  with parameter $\omega_{\mu\nu}$ and supersymmetry
transformations with a sixteen-component Majorana-Weyl spinor Grassmann
parameter, $\epsilon$.    These supersymmetries are those that are  
unbroken by
the presence of a single D-instanton.   The transformations of the   
 variables
that leave the action invariant  are
\begin{eqnarray}\label{sym1}
  \delta A_{A\mu} &=&\epsilon^{ABC} A_{B\mu} \Lambda^C
  +    \omega_{\rho\lambda}(\Sigma^{\rho\lambda})^\nu_{\;\mu}
  A_{A\nu} +  i \bar{\epsilon} \Gamma^\mu\psi_A\\
\delta\psi_A&=&\epsilon^{ABC}\psi_B
  \Lambda^C+\omega_{\rho\lambda}\Gamma^{\rho\lambda}\psi_A+  
\epsilon^{ABC}
A_{B\mu}  A_{C\nu} \Gamma^{\mu\nu}\epsilon\label{sym2}.
\end{eqnarray}
The vector $A_{3\mu}$ will be taken  as the element of the $U(1)$  Cartan
subalgebra which is associated with the relative position of the two
instantons.   This choice breaks the $SU(2)$ symmetry. 

We will see that there are two distinct types of terms that  
contribute to the
partition function so that ,
\begin{equation}
\label{onetwoterm}
{\cal Z}^{(2)} = {\cal Z}^{(2)}_1 + {\cal Z}^{(2)}_2.
\end{equation}
The first term    arises by soaking up all sixteen components of  
$\psi_3$ with
external sources  and has the form,
\begin{equation}
\label{oneterm}
{\cal Z}^{(2)}_1 = \int d^{10} A_3 (J^3)^{16} e^{-V(|A_3|)}
\end{equation}
The second term on the right-hand side of (\ref{twoterm}) is  
proportional to
$(J^3)^8$ and arises because the action, $S$, is independent of half the
$\psi_3$'s.   Since  there is no covariant  way of  eliminating  
half a spinor
we will make use of a light-cone Minkowski space description which  
gives a
result of the form,
\begin{equation}
\label{twoterm}
{\cal Z}^{(2)}_2 = \int d^{10} A_3 (J^3)^8  X(A_3) e^{-V(|A_3|)},
\end{equation}
where  $X(A_3)$ is a function to be determined later.

We  will first  calculate the function $ \exp(-V(|A_3|)$, which involves
setting
 $\psi_{3} =0$   in the action (\ref{dinstaction}).
For the moment we will  treat the problem in arbitrary dimension  
$d$.    For
fixed $A_3$ the action is simplified by choosing coordinate axes so  
that $A_3^d
=L$ and $A_3^i =0$ for $i=1,\cdots, d-1$, so that
\begin{equation}
  \int d^{d}A_{3\mu}=\int dL L^{d-1} d\Omega^{d-1} ,
\end{equation}
where $d\Omega^{d-1}$ is the volume of the unit $(d-1)$-sphere. 
 The dimensionality of the spinor will be denoted by $p$, where
$p=2^{(d-2)/2}$ for minimal spinors in $2,3,4,6$ or $10$  
dimensions.   The
integrations over the internal fermions are very simple,  giving a  
factor of
\begin{equation}
   \int d^{p} \psi_1d^{p}
  \psi_2   \exp( \kappa^{-1}  L\bar{\psi}_2\Gamma^d \psi_1)= \kappa^{-p}
L^{p}.\end{equation}

Now we can  integrate out the internal bosonic variables $A_{1\mu}$ and
$A_{2\mu}$.  Recall that in this discussion these are $SO(d)$  
vectors.  The
bosonic action reduced to zero dimensions is given by,
\begin{eqnarray}\label{bosona}
 S  = {1\over 4\kappa } \left\{ L^2 (A_1)^2+ L^2 (A_{2})^2    +  
(A_1)^2(A_2)^2-
(A_1\cdot A_2)^2  + (a_1A_2 -  a_2 A_1)^2  \right\},
\end{eqnarray}
 where $A_1^i$ and $A_2^i$ are $(d-1)$-vectors and $a_1 = A_1^d$  
and $a_2 =
A_2^d$.  The   $SO(d-1)$ symmetry of this action
  can be used to write the integration measure for the internal
integrations as,
\begin{equation}
  \int d^dA_1 d^dA_2= \int d a_1 da_2 \int
  dR_1 dR_2 d\Omega_1^{d-2}d \Omega_2^{d-3} d\theta R_1^{d-2}  
R_2^{d-2} (\sin
\theta)^{d-3},
\end{equation}
where   $|A_1| = R_1$, $|A_2| = R_2$ and $|A_1\cdot A_2| = R_1 R_2 \cos
\theta$.   The
action  (\ref{bosona}) expressed in these variables is,
\begin{equation}
    S= {1\over 4\kappa} \left\{ L^2\left(
R_1^2+R_2^2\right)+R_1^2R_2^2(\sin\theta)^2+
  a_1^2R_2^2+ a_2^2R_1^2- 2a_1  a_2 R_1R_2\cos\theta\right\}.
\label{sred}
\end{equation}
The gaussian integration over $a_1$ and $a_2$  is simple to  
evaluate, leading
to a factor of $\kappa \pi (R_1R_2 \sin\theta)^{-1}$.
Defining $x = (4\kappa)^{-1} L^2 R_1^2$ and $y= (4\kappa)^{-1} L^2  
R_2^2$, the
partition function  becomes,
\begin{eqnarray}
 e^{-V(L)} &=& \kappa^{d-1-p}   L^{4-2d+ p}  \int dx dy d\theta  (x  
 y\sin^2
\theta)^{(d-4)/2}  e^{\left( -  (x+y) -  4\kappa L^{-4}
xy\sin^2\theta\right)}\nonumber\\
&=& \kappa^{d-1-p}  L^{4-2d+  p}    \int dy d\theta  {y^{(d-4)/2}
(\sin\theta)^{d-4} \over (1 +  4\kappa L^{-4}y\sin^2\theta)^{(d-2)/2}
}e^{-L^2y}.
\label{intz2}
\end{eqnarray}
Here, and in  subsequent  equations,  we   ignore   an overall  
constant factor
in ${\cal Z}^{(2)}$ which is proportional to the volume of the bosonic
integrations and  is independent of $\kappa$ and $L$.

Using (for even $d$)
\begin{equation}
\label{funeq}
  \int_0^\pi d\theta {(\sin\theta)^{(d-4)}\over
  (a^2+b^2\sin^2\theta)^{(d-2)/2}}={(d-3)!!\over 2^{(d-2)/2}\left((d-2)/
2)\right)!} {1\over a(b^2+a^2)^{(d-3)/2}}
\end{equation}
 (see, for example, \cite{gradshteyn} section 3.642 eq.(3)),
it follows that  the potential $V(L)$ is given  by,
\begin{equation}
\label{vldef}
e^{-V(L)} =  \kappa^{(2d-3p)/4} \hat  L^{4- 2d +p}   \int_0^\infty
 dy  {y^{(d-4)/2}\over (1+ 4y \hat L^{-4} )^{(d-3)/2}} e^{-y} ,
\end{equation}
where $\hat L = \kappa^{-1/4} L$.

In the  limit  of large separations, $L\to \infty$, the   potential in
(\ref{vldef}) behaves as,
\begin{equation}
\label{asymptotic}
V(L) \to (2d -4 -p) \ln \hat L + O(\hat L^{-4}).
    \end{equation}
  For pure Yang--Mills theory ($p=0$)  in $d>2$ this is badly  
behaved (and
${\cal Z}^{(2)}_1$ is not extensive) but for the supersymmetric  
Yang--Mills
theories, which have $d=3,4,6$ or $10$ and $p=2,4,8$ and $16$  
respectively, the
potential vanishes asymptotically and the partition function is  
proportional to
the volume, $\int d^{d}A_3$.    Furthermore, in the  
nonsupersymmetric cases a
non-zero metric is generated for the fluctuations $\delta A_{3\mu} \delta
A_{3\nu}$, which arises from the effect of integrating over the internal
bosons.  In the  supersymmetric cases the internal fermion  
integrations cancel
this and the metric remains flat.

The exact expression for the potential shows a  changeover due to  
higher order
$\kappa$ effects at distances of order $\hat L \sim 1$, or $L \sim
\kappa^{1/4}$.  Recalling that the  ten-dimensional Planck distance  
is given,
in string theory, by
\begin{equation}
\label{plancks}
l_P = \kappa^{1/4} \sqrt{\alpha'}
\end{equation}
(the string scale $\sqrt{\alpha'}$ has been set equal to one in the  
forgoing
discussion)
we see that the two-instanton potential develops non-trivial  
dependence at
around the Planck scale, which is the scale at which the string coupling
becomes strong in the classical D-instanton solution.  Precisely  
how this is
reflected in  physical processes is not apparent from this analysis  
but it
seems likely to be of importance in understanding the dynamics of  
string theory
more completely.

The potential, $V(L)$,  is a free energy that encodes the sum over  
all loop
diagrams, which can be recovered by expanding it in a power series in
$\kappa/L^4$.  This can be seen to reproduce the sum over \lq  
Feynman diagrams'
derived from the \lq action',  (\ref{dinstaction}), when  $\psi^3$  
is  soaked
up by external sources.   One simple way of determining the  
coefficients in
this series is to express   (\ref{vldef}) in terms of  the  
parabolic cylinder
function, $D_{3-d}(w) $ (defined in \cite{gradshteyn} section 3.383  
eq.7),
\begin{equation}
\label{ddef}
e^{-V(L)} = (const.) \kappa^{(2d-3p)/4}    w^{(p-2)/2} e^{w^2/4}  
D_{3-d} (w),
\end{equation}
where $w^2 = \hat L^4/2$.   The second-order differential equation  
satisfied by
  $D_n(w)$    (\cite{gradshteyn} 9.255 eq.3) leads to,
  \begin{equation}
\label{morefacts}
{d^2 V  \over dw^2} -   \left({d  V \over dw}\right)^2 - \left( w + (p-2)
{1\over w}\right) { d  V \over dw } - {p(p-2)\over 4 w^2}   - {p +  
4 -2d\over
2}  =0.
\end{equation}
This equation can be solved iteratively in powers of $w^{-2}$,  
giving a  series
that  summarizes the sum over all closed-string world-sheets  with  
boundaries
that are fixed (in all directions) on either of the two  
D-instantons.       As
in the case of the bosonic string  \cite{greentwo}, the logarithm of the
partition function is a sum of {\it connected} world-sheets.
 Importantly, it is only for the supersymmetric cases that the  
constant term is
absent from  (\ref{morefacts}).   In that case $V=0$ as $w\to  
\infty$, which
corresponds to the vanishing of the lowest-order diagram -- the  cylinder
diagram which  vanishes by the abstruse Jacobi identity  due to the
cancellation of the loop of internal bosons and the internal fermions
\cite{greeniib}.    This is the same as the  vanishing of the  
one-loop diagram
in the super-Yang--Mills theory.   In the non-supersymmetric case  
the constant
term in (\ref{morefacts}) leads to a logarithmic divergence in the  
potential
at large $L$ -- correspondingly, the cylinder diagram is non-zero in the
bosonic string theory.    Each successive term adds a boundary  in  
the sum over
world-sheets (which inserts a vertex and two propagators in the Feynman
diagrams for the Yang--Mills theory).  These higher-order  diagrams are
non-vanishing.

 In the classical  field theoretic description  of  \cite{gibbons} the
configuration of  two singly charged wormholes   does not  merge  
smoothly into
a single doubly charged wormhole when the relative separation vanishes.
However, this is the region in which the classical solution is not
 expected to be a good guide.
{}From the integral (\ref{vldef}) it is clear that near the origin,  
$L=0$, the
potential behaves as
\begin{equation}
\label{origv}
V(L) \sim - (p-2) \ln \hat L.
\end{equation}
  This singularity is the signal that the variables $A_3$ and  
$\psi_3$ are not
a complete description of the moduli space.  At the origin the \lq  
stretched
strings' become important as their \lq mass' vanishes  -- in  
particular, the
action becomes independent of the fermionic variables $\psi_1$ and  
$\psi_2$
when $L=0$ so that they are supermoduli, and  integrating over them  
causes $Z^{
(2)}$ to vanish.   This resolution of the apparent  singularity at  
points in
moduli space where instantons coincide is characteristic of the D-brane
picture.

The preceding discussion applies to the $(J^3)^{16}$ term  in which  
all sixteen
components of the spinor, $\psi_{3}$,  are soaked up (from now on  
we will stay
with the ten-dimensional case).  However,  supersymmetry  implies   
that  the
action   depends on eight  components of $\psi_3$  so that half of the
components can be integrated in    (\ref{partfun}) and there are  
only eight
independent  supermoduli, which is half the number of components in a
ten-dimensional Weyl spinor.  This gives the  ${\cal Z}^{(2)}_2$  term in
(\ref{onetwoterm}).   To exhibit this explicitly requires a non-covariant
choice of coordinates which will be motivated here  by the light-cone
description of $D$-instantons in string theory \cite{greeniib,gutgreen}.
There should presumably be a more covariant method of describing the
supermoduli that  involves extra gauge degrees of freedom, such as  
those that
enter in the $\kappa$-symmetry of the manifestly covariant formulation of
superstring theory.

We will begin this part of the  discussion with   lorentzian  
signature and
define light-cone coordinates with respect to two particular  
directions so that
 a $SO(9,1)$ vector decomposes into $SO(8) \times U(1)$,
\begin{eqnarray}
10 &\to & 8_0 \oplus \ 1_1 \oplus   \ 1_{-1}\nonumber\\
A_\mu &\to &  A^i,  \  \  \  A^+, \  \     A^-
\label{vecs}
\end{eqnarray}
where $A^{\pm}=(A^0\pm A_9)/\sqrt{2}$.  The inner product of two  
vectors is
$A_\mu B^\mu=   A_+B^+  +  A_-B^+ +A_iB_i$ ($i=1,\cdots,8$) where   
$A_+= - A^-$
and $A_-= - A^+$.  The $SO(9,1)$ gamma matrices, $\Gamma^\mu$,  satisfy
$(\Gamma^+)^2=(\Gamma^-)^2=0$ and the expressions  
$\Gamma^+\Gamma^-/2$  and
$\Gamma^-\Gamma^+/2$  are  projectors that decompose a  
sixteen-component chiral
 $SO(9,1)$ spinor into  the two inequivalent $SO(8)$ spinors, ${\bf  
8_s}$ and
${\bf 8_c}$.  The two inequivalent  $SO(8)$ spinors which will be  
represented
by undotted and dotted eight-component spinor indices so that  the
supersymmetry moduli   decompose as follows,
\begin{eqnarray}
\label{dotdef}
 10 &\to &   8_{ - \half } \oplus  8_\half\nonumber\\
\psi &\to &  \psi^a , \  \ \ \ \dot\psi^{\dot a}
 \end{eqnarray}
where  $a, \dot a =1,\dots,8$.    From  now on all spinors will be  
dotted or
undotted $SO(8)$ spinors and the  gamma matrices, $\Gamma^\mu$, will be
decomposed into $SO(8)$ matrices, $\gamma^i_{a\dot b}$ and  
$\gamma^i_{\dot a
b}$ (and the index $a$ will from now on take eight values).

We will use the $SU(2)$ symmetry,
 $\delta A_a^+ =\epsilon^{abc} A_b^+ \Lambda^c$,  to rotate to the \lq
light-cone gauge',
\begin{equation}\label{condition1}
  A_1^+ =  0\quad,\quad A_2^+=0.
\end{equation}
This means that the integration variables $A_1^+$ and $A_2^+$ are  
replaced by
$\Lambda^2$ and $\Lambda^1$, with a jacobian factor of
\begin{equation}
\label{jacsu}
\left| {\delta A_1^+ \over \delta \Lambda^2} {\delta A_2^+ \over \delta
\Lambda^1}  \right| = (A_3^+)^2.
\end{equation}

We now wish to transform away other variables by making use of  
those $SO(9,1)$
and supersymmetry transformations which do not affect the condition
(\ref{condition1}).  We first consider the supersymmetry transformations,
\begin{eqnarray}  \delta\dot \psi_3 &=&  A_1^i A_2^j  \gamma^{ij}  
\dot \rho +
(A_1^- A_2^i  - A_2^- A_1^i ) \gamma^i  \rho, \label{frotwo}\\
 \delta \psi_3 &=&   A_1^i A_2^j \gamma^{ij} \rho.
\label{frotwob}
\end{eqnarray}
The light-cone gauge conditions (\ref{condition1}) are spoilt by the
transformations associated with the $\rho$ components but the $\dot \rho$
components  can be used to  eliminate  the dotted components by setting
\begin{equation}\label{condition2}
 \dot \psi_3^{\dot{a}}=0.
\end{equation}
This replaces the $\psi_3^{\dot{a}}$ integrations with integrations  
over the
components of $\dot \rho^{\dot a}$ with a jacobian,
\begin{equation}\label{fermjacobian}
\left|{\partial\dot \psi^{\dot a} \over \partial\dot \rho^{\dot  
b}}\right| =
{1\over Pf(A_1^i A_2^j\gamma^{ij}_{\dot a\dot b})},
\end{equation}
where $Pf$ denotes the Pfaffian, which can easily be evaluated  
explicitly to
give,
\begin{equation}
\label{paffdef}
Pf(A_1^i A_2^j\gamma^{ij}_{\dot a\dot b}) = {9\over 2} \left[  
(A_1\cdot A_2)^2
- (A_1)^2 (A_2)^2\right]^2.
\end{equation}

Finally,  we can make use of the subset of the  $SO(9,1)$  
transformations,
\begin{equation}\label{trafo3}
  \delta A_3^i = \omega_{ij}A_3^j + \omega_{i-}A_3^-  +  \omega_{i+}A_3^+
\end{equation}
that do not spoil either of the conditions,
(\ref{condition1}) or (\ref{condition2}).   The $\omega_{ij}$  
transform of
$\dot \psi_3^{\dot a}$,  violates (\ref{condition2})  and  the  
$\omega_{i -}$
transform of  $A^{i}$ violates (\ref{condition1}),   but   the  
$\omega_{i+}$
boosts can be used  to impose the condition,
\begin{equation}\label{condition3}
  A_3^i =0.
\end{equation}
This replaces the variables  $A_3^i$ by $\omega^{i -}$, introducing  
a jacobian,
\begin{equation}
\label{rotjacob}
\det {\partial A_3^i  \over  \partial \omega^{j -}} =(A_3^+ )^8.
\end{equation}

When expressed in terms of the coordinates implied by the conditions
(\ref{condition1}), (\ref{condition2}), (\ref{condition3}) the  
action becomes,
$S' = S_b + S_f$, where,
\begin{eqnarray}
 S_b&=& {1\over 4 \kappa}\left\{   -  A_3^-  A_3^+ \left( (A_1 )^2+(  A_2
)^2\right) + (A_3^+)^2 \left((A_1^-)^2+(A_2^- )^2\right)  
\right.\nonumber\\
 &&\left. (  A_1)^2 (  A_2)^2  -(  A_1\cdot   A_2  
)^2\right\}\label{bosmas}\\
S_f&=& {1\over \kappa}
\left\{A_2^i {\psi}^{1\dot{a}}\gamma^i_{\dot{a}a}\psi_3^a +A_2^-  
{\psi}_1^a
\psi_3^a  \right.\nonumber\\
&&\left.
 - A_1^i \dot \psi_2^{\dot{a}}\gamma^i_{\dot{a}a}\psi^{3a} - A_1^-  
\psi_2^a
\psi_3^a  + A_3^-  \psi_1^a  \psi_2^a +  A_3^+ \dot  
\psi_1^{\dot{a}}   \dot
\psi_2^{\dot{a}}\right\}.
\label{fermas}
\end{eqnarray}
The partition function, ${\cal Z}^{(2)  }$, is now given  by (again
dropping a $\kappa$ independent constant)
\begin{equation}
\label{newpart}
{\cal Z}^{(2)}_2 = \int (J^3)^8  d^8 \psi_3^a  d^8\omega^{i-}
dA_3^+ dA_3^-  { (A_3^+)^{10}\over Pf(A_1^iA_2^j \gamma^{ij})}    
Z^{\prime } ,
\end{equation}
where $( J^3)^8$ soaks up the $\dot \rho$ integrations,  and
$Z^{\prime }$ is defined by  the internal integrations,
\begin{equation}
\label{lavs}
Z^{\prime} =\int d^8A_1^i d^8A_2^i  d^8\dot\psi_1^{\dot a}  
d^8\dot\psi_2^{\dot
a} d^8 \psi_1^a d^8 \psi_2^a  \exp(-S').
 \end{equation}

The fermionic fields can be integrated out  by first redefining  
$\psi_1$  and
$\psi_2$ by,
\begin{eqnarray}\label{shift1}
 && \psi_1^a\to \psi_1^a - {A_1^- \over A_3^-}\psi_3^a ,\qquad\  
\quad  \psi_2^a
\to \psi_2^a - {A_2^- \over A_3^- }\psi_3^a \\
&&\dot \psi_1^{\dot{a}} \to \dot  \psi_1^{\dot{a}} - {A_1^i \over A_3^+
}\gamma^i_{\dot{a}a} \psi_3^a ,  \qquad \dot \psi_2^{\dot{a}}\to \dot
\psi_2^{\dot{a}} - {A_2^i \over A_3^+} \gamma^i_{\dot{a}a}\psi_3^a
\label{shift2} .
\end{eqnarray}
This completes the square on $\psi_3^a$ so that  substituting
(\ref{shift1}), (\ref{shift2}) into (\ref{fermas}) gives,
\begin{equation}
 S_f={1\over \kappa}\left\{
 A_3^-  \psi_1^a \psi_2^a +  A_3^+    \dot \psi_1^{\dot{a}}  \dot
\psi_2^{\dot{b}} - {A_1^i A_2^j \over A_3^+ } \psi_3^a  \gamma^{ij}_{ab}
\psi_3^b\right\}.
\label{finferm}
\end{equation}
Integration over $\psi_1$ and $\psi_2$ gives a factor of $ \kappa^{-16}(A_3^-  
A_3^+)^8$ in
the measure.

 At this stage we could soak up the $\psi_3^a$'s with  a source  
term, which is
precisely the situation we discussed earlier and the resulting bosonic
integrations lead to (\ref{oneterm}) after a Wick rotation.   
However, we see
that  the integration over $\psi_3^a$ is non-zero -- in fact, from  
the last
term in (\ref{finferm}) the  integration over these components    
gives a factor
$(\kappa A_3^+)^{-4} Pf(A_1^i A_2^j \gamma^{ij})$, cancelling the Pfaffian in
(\ref{newpart}).     This eliminates all the fermionic fields from  
the action,
leaving the eight components of $\dot \rho^{\dot a}$ as the only  
fermionic
integration variables.  At this point the bosonic integrations bear  
a close
similarity to the Minkowski signature version of the integrations  
carried out
earlier, but with a different parameterization of the internal  
variables, $A_1$
and $A_2$.    The integral over $A_1^-$ and $A_2^-$ is Gaussian and  
gives a
factor of  $\kappa  (A_3^+)^{-2}$ so that the   integral (\ref{newpart})
reduces to
\begin{equation}\label{intS}
{\cal Z}^{(2) }_2=\int   (\dot J_3)^8   d^8  
\omega^{i-} \; dA_3^+
dA_3^- \;(A_3^+ )^{12}(A_3^- )^8 \Sigma ,
\end{equation}
where
\begin{equation}
\label{zprimes}
\Sigma = \int d^8A_1^i d^8 A_2^i \;\exp(-S_b)
\end{equation}

Before  integration over the remaining internal coordinates   
$A_1^i$ and $
A_2^i$ we want to make contact with the euclidean theory   by replacing
$A_3^0$ by $iA_3^{10}$, which has the effect of replacing $A_3^\pm$ by
 \begin{equation}
\label{adef}
A= {1\over \sqrt 2} (A_3^9 + iA_3^{10}), \qquad \bar A =   {1\over  
\sqrt 2}
(A_3^9 - i A_3^{10}) .
\end{equation}
  With these conventions we make the replacement,
\begin{equation}
\label{euclids}
 - A_3^- A_3^+  \to   (A_3^9)^2 + (A_3^{10})^2 \equiv L^2 = A \bar A.
\end{equation}
The  $\omega^{i-}$ variables
are now  identified with the  rotation generators that transform  
the commuting
$SO(2)$ and $SO(8)$ subgroups of $SO(10)$ into each other.

The integration over   $A_1^i$ and $A_2^i$ can now be carried out by
transforming to eight-dimensional  polar coordinates,
\begin{equation}
\label{eightpolar}
A_1^i  = R_1 n_1^i, \qquad A_2^i = R_2 n_2^i,
\end{equation}
where $n_1^2 = n_2^2 =1$ and $n_1 \cdot n_2 = \cos \theta$.
\begin{equation}\label{ameasure}
 \Sigma= \int d\Omega^7 d\Omega^6\int dR_1 dR_2d\theta  R_1^7R_2^7(\sin\theta)^6
\;\exp(- S_b) ,
\end{equation}
where  the action is
 \begin{equation}
S_b ={1\over 4\kappa}\left\{  L^2 (R_1^2+R_2^2)  +
 (R_1)^2(R_2)^2\sin^2\theta  \right\}.
\end{equation}
The evaluation of the $R_1$, $R_2$ and $\theta$ integrals follows  
closely the
steps from (\ref{sred}) to (\ref{vldef}), leading to $\Sigma (L) =  
V(|L|)$.
 The  two-instanton partition function (\ref{intS}) then reduces to
 \begin{equation}
\label{partsumm}
{\cal Z}^{(2)}_2 = \kappa^{-11} \;\int    \; dA d\bar A\; d^8 A_3^i\;
  A^{-4}  \;
e^{-V(|A_3|)} (J_3)^8,
\end{equation}
which is of the form (\ref{twoterm}).    Although this expression is not
manifestly $SO(10)$ invariant it should  give rise to covariant  
scattering
amplitudes (since the external sources  that soak up the eight   
fermionic zero
modes  also have a noncovariant form in the light-cone gauge).   

This expression again has an obvious translation into the string theory
D-instanton free energy, which is obtained by summing over closed  
world-sheets
with boundaries on the two D-instantons.  Contact is made with the  
light-cone
parameterization  by  Fourier transforming with respect to $Y^- = Y^-_1 -
Y^-_2$ and transforming to the light-cone parameterization of
\cite{mandelstam} in which the surfaces are flat apart from the  
interaction
points at which closed strings split and join.    Each term  in  
this series has
the topology of a closed-string diagram, but with the initial and final
closed-string states located at the finite times $Y_1^+$ and $Y_2^+$.
  The leading term in this sum  --  the cylinder diagram  --   is  
non-zero when
eight undotted fermionic open-string states are attached to the  
boundaries.
These are supplied by the    $(J^3)^8$ term.    Attaching these  
eight fermions
is analogous  to picking out the (velocity)$^4$ term in considering  
the force
between D-particles in \cite{bachas,pouliot,sodeberg,shenker}.     The
expression  for this  cylinder contribution  is   given in
\cite{greeniib,gutperle}.   Only the massless closed-string states  
contribute
-- just as in the D-particle case -- and the result is proportional  
to $(p^+)^4
L^{-8}$, where $p^+ \sim {\partial/ \partial L}$, and the result  
behaves as
$L^{-12}$.    The same result is obtained by viewing the process in  
the annulus
channel, which is a trace over states of open strings with fixed  
end-points.
Only the lowest states of the stretched strings contribute here also  --
precisely the same supermultiplet that enters into the Yang--Mills  
action.  The
leading diagram is the one-loop eight-point function with external  
$\psi_3^a$'s
which have vertices of the form $A_1^i \psi_2^{\dot a}  
\gamma^i_{\dot a a}
\psi_3^a -  A_2^i \psi_1^{\dot a} \gamma^i_{\dot a a} \psi_3^a$.     The
$A_{1,2}^i$ propagators are $1/L^2$ while the $\psi_{1,2}^{ \dot a}$
propagators are $1/L$, so the diagram is proportional to $1/L^{12}$ in
agreement with the cylinder interpretation.

The higher-order  terms in the sum over the Yang--Mills diagrams  
translate into
world-sheets which contain both planar and non-planar contributions  
-- the
latter encoding the effects of handles, or gravitational  
corrections.  Although
the light-cone treatment does not easily generalize to $N>2$,  the string
world-sheet calculation has an obvious covariant formulation. It  should
therefore be possible to understand the counting of fermionic modes  
for general
$N$ in a more covariant manner -- possibly by embedding the system  
in one with
more local symmetry, such as the $\kappa$ symmetry.  The limit of  
large $N$,
where  closed-string loop corrections are suppressed if $\kappa  
N^2$ is fixed,
might be particularly interesting to analyze.

Finally, it is salutory to remember that the Yang--Mills  
approximation to the
configuration space does not include the winding modes of Dirichlet  open
strings.  These must play an important r\^ole in the  
compactification to nine
dimensions on a circle of radius $R$.  The Dirichlet open-string   
has winding
numbers which  transform under T-duality into momenta spaced by $\sqrt
{\alpha'}/R$.   The instanton then  describes the euclidean  
compactification of
the type IIA D-particle world-line.     In the  limit  $R \to 0$, the
one-dimensional  Yang--Mills theory describing    ten-dimensional  
D-particles
is recovered.   Clearly, the considerations of this paper must be  
generalized
at scales $L \le R$ to include the open-string winding modes, which are
discrete momenta for the super Yang--Mills theory.   This should  
lead to a
cross-over between effects that arise at the ten-dimensional Planck  
scale in
the type IIB theory and the eleven-dimensional Planck scale in type IIA,
analogous to the cross-over  described in \cite{shenker}.

\vskip 0.8cm
\noindent{\it Acknowledgments}:

\noindent  We are grateful to Michael Douglas, Gary Gibbons,  
Malcolm Perry and
Chris Hull for useful conversations.

\end{document}